\documentclass[preprint,preprintnumbers,amsmath,amssymb]{revtex4}

\usepackage{graphicx}% Include figure files
\usepackage{dcolumn}% Align table columns on decimal point
\usepackage{bm}% bold math

\newcommand{\mb}[1]{ \mbox{\boldmath$#1$} }
\newcommand{\ds}{\displaystyle}
\newcommand{\x}{\mbox{\boldmath$x$}}
\newcommand{\y}{\mbox{\boldmath$y$}}
\newcommand{\beq}{\begin{eqnarray}}
\newcommand{\eeq}{\end{eqnarray}}
\newcommand{\beqq}{\begin{eqnarray*}}
\newcommand{\eeqq}{\end{eqnarray*}}

\begin{document}

\title{ATTENUATION OF THE ELECTRIC POTENTIAL AND FIELD IN DISORDERED SYSTEMS\\}
\author{A. Singer}
\email{amits@post.tau.ac.il}
\author{Z. Schuss}
\email{schuss@post.tau.ac.il}
\affiliation{Department of Applied Mathematics\\
Tel-Aviv University, Ramat-Aviv\\
69978 Tel-Aviv, Israel}
\author{R. S. Eisenberg}
\email{beisenbe@rush.edu} \affiliation{Department of Molecular
Biophysics and Physiology\\ Rush Medical Center, \\ 1750 Harrison
St., Chicago, IL 60612\\}

\begin{abstract}
We study the electric potential and field produced by disordered
distributions of charge to see why clumps of charge do not produce
large potentials or fields. The question is answered by evaluating
the probability distribution of the electric potential and field in
a totally disordered system that is overall electroneutral. An
infinite system of point charges is called totally disordered if the
locations of the points and the values of the charges are random. It
is called electroneutral if the mean charge is zero. In one
dimension, we show that the electric field is always small, of the
order of the field of a single charge, and the spatial variations in
potential are what can be produced by a single charge. In two and
three dimensions, the electric field in similarly disordered
electroneutral systems is usually small, with small variations.
Interestingly, in two and three dimensional systems, the electric
potential is usually very large, even though the electric field is
not: large amounts of energy are needed to put together a typical
disordered configuration of charges in two and three dimensions, but
not in one dimension. If the system is locally electroneutral---as
well as globally electroneutral---the potential is usually small in
all dimensions. The properties considered here arise from the
superposition of electric fields of quasi-static distributions of
charge, as in nonmetallic solids or ionic solutions. These
properties are found in distributions of charge far from
equilibrium.
\end{abstract} \maketitle

\section{Introduction}
There is no danger of electric shock when handling a powder of salt
or when dipping a finger in a salt solution, although these systems
have huge numbers of positive and negative charges. It seems
intuitively obvious that the alternating arrangement of charge in
crystalline Na$^+$Cl$^-$ should produce electric fields that add
almost to zero; it also seems obvious that Na$^+$ and Cl$^-$ ions
will move in solution to minimize their equilibrium free energy and
produce small electrical potentials. But what about random
arrangements of charge that occur in a random quasi-static
arrangement of charge such as a snapshot of the location of ions in
a solution? Tiny imbalances in charge distribution produce large
potentials, so why doesn't a random distribution of charge produce
large potentials, particularly if the distribution is not at
thermodynamic equilibrium? Indeed, some arrangements of charge
produce arbitrarily large potentials, but as we shall see, these
distributions occur rarely enough that the mean and variance of
stochastic distributions are usually finite and small. More
specifically, we determine the conditions under which stochastic
distributions of fixed charge produce small fields.

The quasi-static arrangements of charge can represent the fixed
charge in amorphous non-metallic solids or snapshots of charge
arrangement of ions in solution, due to their random (Brownian)
motion. Our analysis does not apply to quantum systems
\cite{Brydges}, and in particular it fails if electrons move in
delocalized orbitals, as in metals. Note that the random
arrangements of charge considered here do not necessarily minimize
free energy.

We consider the field and potential in overall electroneutral random
configurations of infinitely many point charges. An infinite system
of point charges is called totally disordered if the locations of
the points and the charges are random, and it is called overall
electroneutral if the mean charge is zero. The configurations of
charge may be static or quasi-static, that is, time dependent, but
varying sufficiently slowly to avoid electromagnetic phenomena: the
electric potential is described by Coulomb's law alone. In one
dimensional systems of this type, the potential is usually
finite---even though the system usually contains an infinite number
of  positive and negative charges. Even if the system is disordered
and spatially random, charges of the same sign do not clump together
often enough to  produce large fields or potentials, in one
dimensional systems.

Our approach is stochastic. We ask how disordered can a random
electroneutral system be, yet still have a small field or potential.
We find the answer by evaluating the probability distribution of the
electric potential and field of a disordered system of charges. We
find that the electric field in a totally disordered one dimensional
system is small whether the system is locally electroneutral or not.
The potential behaves differently; it can be arbitrarily large in a
one dimensional system, but it is usually small in electroneutral
systems.

In two or three dimensional disordered systems, the electric field
is not necessarily small. We show that in such systems that are also
electroneutral the field is usually small. The potential, however,
is usually large, even if the system is electroneutral. Both
potential and field are small, if the system is locally---as well as
globally---electroneutral (see definition below) in one, two and
three dimensions.

 We consider several types of random arrays of charges: (a)
A lattice with random distances between two nearest charges; (b) A
lattice (of random or periodic structure) with a random
distribution of positive and negative charges (charge $\pm1$).
Charges in the lattice need not alternate between positive and
negative, nor need they be periodically distributed; (c) A lattice
(of random or periodic structure) with random charge strengths.
Not all charges are $\pm1$, but they are chosen from a set
$q_1,q_2,\dots,q_n$ with probabilities $p_1,p_2,\dots,p_n$,
respectively, such that
 \beq
 \ds\sum_{i=1}^nq_ip_i=0.\label{electroneutral}
 \eeq
Equation (\ref{electroneutral}) is our definition of
electroneutrality in an infinite system.

We use renewal theory \cite{Karlin}, perturbation theory
\cite{Bender}, and saddle point approximation \cite{Jensen} to
calculate the electric potential of one dimensional systems of
charges and show that it is usually small. That is to say, the
probability is small that the potential takes on large values. Thus,
randomly distributed particles produce small potentials even in
disordered systems in one dimension, if the system is
electroneutral. The analysis of one dimensional systems requires the
calculation of the probability density function (pdf) of weighted
independent identically distributed (i.i.d.) sums of random
variables. This pdf looks like the normal distribution near its
center, but the tail distribution has the double exponential decay
of the log-Weibull distribution \cite{Weibull}. We conclude that the
electric potential of totally disordered electroneutral one
dimensional systems is necessarily small, comparable to that of a
single charge.

Later in the paper, we define local electroneutrality precisely
and show that two and three dimensional systems with local
electroneutrality usually have small potentials, because the
potential of a locally neutral system of charges decays like the
potential of a point dipole, as $1/r^2$. We show that the
potential of typical totally disordered arrays of charges in two
and three dimensions is infinite even if the system is
electroneutral.

Historically, little attention seems to have been paid to
quasi-static random arrangements of charge,\textbf{ although much
attention has been paid to the equilibrium arrangements of mobile
charge.} In systems of mobile charges, such as liquids and ionic
solutions, the decay of the electric potential may even be
exponential, after the mobile charges assume their equilibrium
distribution. The early theory of Debye-H\"{u}ckel \cite{Barthel}
shows a nearly exponential decay (with distance from a given
particle) of the average electric potential at equilibrium,
originally found by solving the linearized Poisson-Boltzmann
equation. In classical physics, perfect screening of multipoles
(of all orders) occurs in both homogeneous and inhomogeneous
systems at equilibrium in the thermodynamic limit, when boundary
conditions at infinity are chosen to have no effect
\cite{McQuarrie} and there is no flux of any species. This type of
screening in electrolytic solutions is produced by the equilibrium
configuration of the mobile charges \cite{Henderson,Martin}, which
typically takes 100 psec to establish (compared to the $10^{-16}$
time scale of most atomic motions) \cite{Barthel2}. Many other
systems are screened by mobile charges after they assume their
equilibrium configuration of lowest free energy \cite{Chazalviel},
such as \textbf{ionic solutions,} metals and semiconductors.

The spatial decay of potential in ionic solutions determines many
of the properties of ionic solutions and is a striking example of
screening or shielding. ``Sum rules" of statistical mechanics
\cite{Henderson,Martin} describe these properties. These rules
depend on the system assuming an equilibrium distribution, which
can only happen if the charges are mobile.

We consider finite and infinite systems of charges \textbf{which
may or may not be mobile and which are not necessarily at
equilibrium.} We show that the potential of a finite disordered
locally electroneutral system is attenuated to the potential of a
single typical charge, whether the potential is evaluated inside
or outside a finite system or in an infinite system. We note that
the behavior of the electric potential and field outside the line
or plane of the lattice can be analyzed in a straightforward
manner by the methods developed below.

\section{A one dimensional ionic lattice}
\begin{figure}
\includegraphics{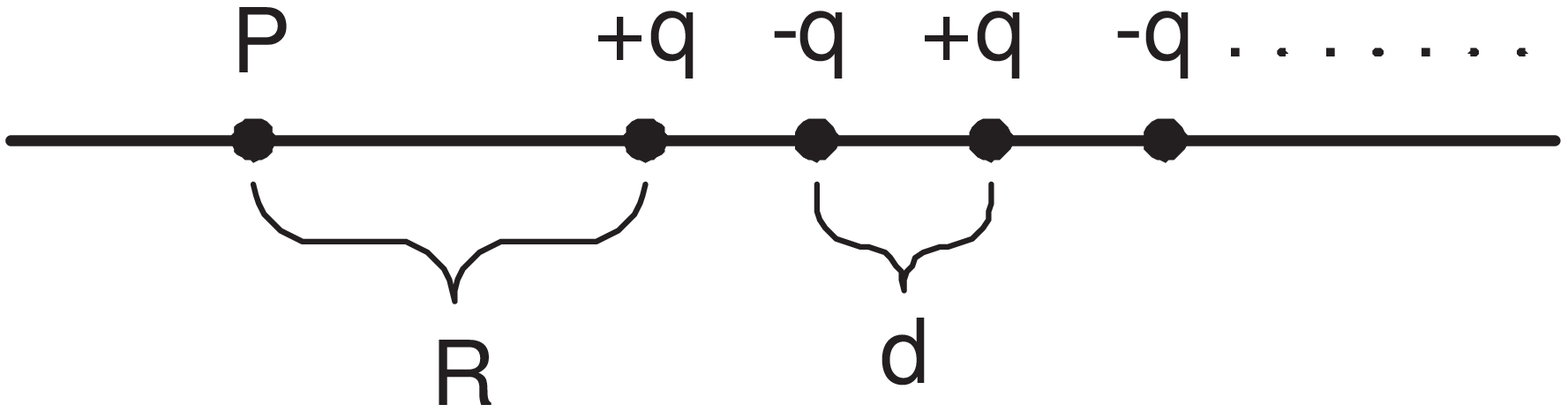}
\caption{A semi infinite lattice of alternating charges with a
distance $d$ between neighboring charges. The point $P$ is located
at a distance $R$ from and to the left of the first
charge.}\label{f:1D}
\end{figure}

Consider a semi-infinite array of alternating electric charges
$\pm q$ with a distance $d$ between neighboring charges. The
electric potential $\Phi$ at a point $P$, located at a distance
$R$ from and to the left of the first charge (see Fig. \ref{f:1D})
is given by
\begin{eqnarray}
\Phi &=& \frac{q}{4\pi \varepsilon_0}
\left(\frac{1}{R}-\frac{1}{R+d}+\frac{1}{R+2d}-\frac{1}{R+3d} +
\cdots \right) \nonumber \\
& = & \frac{q}{4\pi \varepsilon_0 R} \left(1-\frac{1}{1+a} +
\frac{1}{1+2a}-\frac{1}{1+3a}+\cdots \right) \nonumber \\
& = & \frac{q}{4\pi \varepsilon_0 R} \sum_{n=0}^\infty
\frac{(-1)^n}{1+na}, \label{series}
\end{eqnarray}
where $a=d/R$ is a dimensionless parameter. The series
(\ref{series}) is conditionally convergent, so it can be summed to
any value by changing the order of summation \cite{Knopp}.
\textbf{The order of summation reflects the order of construction
of the system; different orders may lead to different potential
energies of the system. However, the infinite series that
determine the electric field are absolutely convergent (see
below), so the field does not depend on the order of summation of
its defining series. Thus, all potentials differ from each other
by a constant, which presumably reflects the different ways the
charge distribution could be constructed, while having the same
electric field. From here on, we consider the ordering in equation
(\ref{series}).}

Setting $R=d$ ($a=1$) we find the potential at a vacant lattice
point (to avoid infinite potentials) due to charges located at
both directions of the infinite lattice is
 $$2\Phi(R=d) = 2\frac{q}{4\pi \varepsilon_0 d} \sum_{n=1}^\infty
\frac{(-1)^{n-1}}{n}=  \frac{q}{4\pi \varepsilon_0 d}\cdot 2\log
 2. $$
The constant $2\log 2$ is known as the Madelung constant of a one
dimensional lattice \cite{Kittel}.

Next we find the asymptotic behavior of the potential $\Phi$ away
from the semi-infinite lattice, that is for $R \gg d$, or
equivalently $a \ll 1$. The following analysis is independent of
the order of summation of the series (\ref{series}). Clearly, the
infinite sum in eq. (\ref{series}) converges, because it is an
alternating sum with a decaying general term. We expand the
potential for $a \ll 1$ (away from the lattice) in the asymptotic
form
\begin{equation}
\label{eq:potential1} \Phi =
\frac{q}{4\pi\varepsilon_0}\frac{1}{R}\left(V_0 + aV_1 + a^2V_2 +
\cdots\right).
\end{equation}
The effect of the first charge can be separated from all the
others,
\begin{equation}
\label{eq:potential2} \Phi =
\frac{q}{4\pi\varepsilon_0}\frac{1}{R} -
\frac{q}{4\pi\varepsilon_0}\frac{1}{R+d}\left(V_0 + \tilde{a}V_1 +
\tilde{a}^2V_2 + \cdots\right),
\end{equation}
where
 \[\tilde{a}=\frac{d}{R+d}=\frac{a}{1+a}.\]
Comparing eqs.(\ref{eq:potential1}) and (\ref{eq:potential2}) we
obtain
 \beqq
V_0 + aV_1 + a^2V_2 + \cdots = 1 -
\frac{1}{1+a}\left[V_0+\frac{a}{1+a}V_1+\left(\frac{a}{1+a}\right)^2V_2
+ \ldots \right].
 \eeqq
The coefficients $V_0, V_1, \ldots$ are found by equating the
coefficients of like powers of $a$. In particular, we find that
$V_0=1/2, V_1=1/4, V_2=0$, so the potential has the asymptotic
form
\begin{equation}
\label{eq:asym} \Phi =
\frac{q}{4\pi\varepsilon_0}\frac{1}{R}\left[\frac{1}{2} +
\frac{1}{4}\,a + O\left(a^3\right) \right].
\end{equation}
All coefficients $V_n$ can  easily be computed in a similar fashion.
This result also determines the rate at which the potential far away
reaches its limiting value, $\ds
\frac{1}{2}\frac{q}{4\pi\varepsilon_0R}$. The divergent series for
$x=1$ has the value $V_0=\frac {1}{2}$ if interpreted as a limit
using the Abel sum \cite{Knopp}
$$1-1+1-1+1-1+\ldots=\lim_{x\to 1^-} \sum_{n=0}^\infty (-1)^n x^n = \lim_{x\to 1^-}
 \frac{1}{1+x} = \frac12.$$
We note that the asymptotic expansion (\ref{eq:asym}) can also be
found directly from the differential equation that the sum
$$y(x)=\sum_{n=0}^\infty \frac{(-1)^n}{1+na}x^n$$
satisfies \cite{Mohazzabi}
\begin{equation}
\label{eq:diff} axy'+y=\frac{1}{1+x},
\end{equation}
with initial condition $y(0)=1$. The asymptotic form of $y(x)$ can
 easily be found by standard methods \cite{Bender}. In particular,
$$\lim_{x\to1-}y(x)=\sum_{n=0}^\infty \frac{(-1)^n}{1+na}.$$

The physical interpretation of the asymptotic expansion
(\ref{eq:asym}) is that the electric potential away from an
infinite lattice of charged particles is about the same as if half
a single charge were located at the origin. The spatial
arrangement of the lattice attenuates the effect of its charge.
The potential near the lattice is determined by a few of the
nearest charges and the contribution of the remaining charges
reduces to that of a half charge placed at a distance $R\gg d$.
Obviously, as $R\to0$ the potential becomes infinite, approaching
the potential produced by just the nearest charge.

\section{One dimensional random ionic lattice}
We turn now to solids in which the charges are distributed
randomly in several different ways. First, consider a
semi-infinite lattice of electric charges, in which the sign of
each charge is determined randomly by a flip of a fair coin. That
is, the charges that are located at the lattice points $X_n
\;(n=0,1,2,\ldots)$ are independent Bernoulli random variables
that take the values $\pm 1$ with probability $1/2$. The electric
potential of this random lattice is given by
\begin{equation}
\label{eq:ber} \Phi = \frac{q}{4\pi \varepsilon_0 R}
\sum_{n=0}^\infty \frac{X_n}{1+na}.
\end{equation}
Some discussion of the nature of convergence of the series
(\ref{eq:ber}) is needed at this point. The convergence of the sum
of variances means that the partial sums converge in $L^2 $ with
respect to the probability measure, so the sum (\ref{eq:ber})
exists as a random variable  $\Phi\in L^2$, whose variance is the
sum of the variances. Now, the Cauchy-Schwarz inequality implies
that $\Phi\in L^1$, so $\langle\Phi\rangle=0$. Note that
(\ref{eq:ber}) also converges with probability 1 \cite{Breiman}.

We use fair coin tossing to maintain the condition of global
electroneutrality, though arbitrary long runs of positive or
negative charges occur in this distribution. Thus some
realizations of the sequence $X_n$ have runs (`clumps') of
substantial net charge and potential. The standard deviation of
the net charge in a region gives some feel for the size of the
clumps. The standard deviation in the net charge of a region
containing $N$ charges is $q\sqrt{N}$. For large values of $N$,
substantial regions are not charge neutral. The condition of local
charge neutrality (defined later) is violated for many of the
realizations of charge in this distribution.

Note that a particular set of $X_n$ can produce an infinite
potential, despite our general conclusions. If, for example,
$X_n=1$ for all $n$, the electric potential becomes infinite (see
eq. (\ref{eq:ber})), because $\ds\sum_{n=0}^\infty
\ds\frac{1}{1+na}=\infty$. Nonetheless, the $L^2$ convergence of
(\ref{eq:ber}) implies that the probability that (\ref{eq:ber}) is
infinite is 0. In other words, even though the potential is
infinite for a particular  set of $X_n$, the potential is finite
with probability 1. This is a striking example of the attenuation
of the electric field, even without mobile charge. The attenuation
of the potential produced by some `clumpy' configurations of
charges occurs even though \emph{there is no correlation in
position, and there is no motion} whatsoever.

The electric field, given by
 \beqq
E = -\frac{q}{4\pi \varepsilon_0 R^2} \sum_{n=0}^\infty
\frac{X_n}{(1+na)^2},
 \eeqq
remains finite for all realizations of $X_n$, because the sum
$$S=\frac{q}{4\pi \varepsilon_0 R^2} \sum_{n=0}^\infty \frac{1}{(1+na)^2}$$ converges.

The electric field is bounded (above and below) by $S$ and so
there is zero probability that the function is outside the
interval $(-S,S)$. The pdf of the electric field is compactly
supported, even when all charges are positive (or negative). The
electric field---unlike the potential---is attenuated even if the
net charge of the system is not zero, taken as a whole. The
standard deviation of the field is $\ds\frac{q}{4\pi \varepsilon_0
R^2}\left\{\sum_{n=0}^\infty \frac{1}{(1+na)^4}\right\}^{1/2}$,
which is of the order of the field of a single charge at a
distance $R$.

\subsection{Moments}
The expected value of $\Phi$ is $\langle\Phi\rangle = 0$, as
mentioned above. The variance of $\Phi$ is given by
\begin{equation}
\label{eq:var} Var{\thinspace}(\Phi) = \left(\frac{q}{4\pi
\varepsilon_0 R}\right)^2 \sum_{n=0}^\infty \frac{1}{(1+na)^2}.
\end{equation}
A vacant lattice point in an infinite (not semi-infinite) lattice
corresponds to $R=d$ for both the charges to the right and to the
left. It follows that the variance of the potential there is twice
that given in (\ref{eq:var}) with $a=1$, that is,
\begin{equation}
Var{\thinspace}(\Phi) = 2\left(\frac{q}{4\pi \varepsilon_0
d}\right)^2 \sum_{n=1}^\infty \frac{1}{n^2} = 2\left(\frac{q}{4\pi
\varepsilon_0 d}\right)^2 \frac{\pi^2}{6},
\end{equation}
so that the standard deviation is
\begin{equation}
\sigma_{\Phi} = \frac{q}{4\pi \varepsilon_0 d}
\frac{\pi}{\sqrt{3}}.
\end{equation}
As expected, the constant $\pi/\sqrt{3}$ is larger than the
Madelung constant $2\log 2$ of the periodic lattice, because the
potential of the disordered system is larger than that of the
ordered one.

Away from the semi infinite lattice, i.e., for $a\ll 1$, we can
approximate the variance (\ref{eq:var}) by the Euler-Maclaurin
formula, which replaces the sum by an integral,
\begin{equation}
Var{\thinspace}(\Phi) = \left(\frac{q}{4\pi \varepsilon_0
R}\right)^2 \left(\int_0^\infty \frac{1}{(1+ax)^2}\,dx +
\frac{1}{2} + O(a)\right) = \left(\frac{q}{4\pi \varepsilon_0
R}\right)^2 \left(\frac{1}{a}+\frac{1}{2}+O(a) \right),
\end{equation}
so the standard deviation is
\begin{equation}
\label{eq:sd} \left.\sigma_{\phi}\right|_R =
\frac{q}{4\pi\varepsilon_0 \sqrt{dR}}\left(1+O(a) \right).
\end{equation}
The decay law of $\ds\frac{1}{\sqrt{R}}$ is more gradual than the
decay law $\ds\frac{1}{R}$ of a single charge.

\subsection{The electrical potential as a weighted i.i.d. sum }
\label{sec:weighted} The potential (\ref{eq:ber}) is a weighted
sum of the form $\sum a_n X_n$, where $X_n$ are i.i.d. random
variables. The distribution of potential is generally not normal.
For example, consider the weighted sum $\sum_{n=1}^\infty
2^{-n}X_n$, where $X_n$ are the same Bernoulli random variables.
This weighted sum represents the uniform distribution in the
interval $[-1,1]$. It is, in fact equivalent to the binary
representation of real numbers in the interval. Not only does this
distribution not look like the Gaussian distribution for small
deviations, it does not look at all Gaussian for large deviations.
In fact, this distribution has compact support. It is zero outside
a finite interval, without the tails of the better endowed
Gaussian. Other unusual limit distributions can be easily obtained
from sums of the form (\ref{eq:ber}). For example, the weighted
sum $\sum_{n=1}^\infty 3^{-n}X_n$ is equivalent to the uniform
distribution on the Cantor ``middle thirds'' set \cite{Cantor} in
$[-1,1]$, whose Lebesgue measure (length) is 0.

Note that the sum
$$\sum_{n=0}^\infty \frac{X_n}{(1+na)^{1+\varepsilon}}$$
has compact support for every $\varepsilon>0$, because the series
$$\sum_{n=0}^\infty \frac{1}{(1+na)^{1+\varepsilon}}$$
converges for every $\varepsilon>0$. In our case $\varepsilon=0$,
so that the limit distribution does not necessarily have compact
support. Nonetheless, we expect that the probability distribution
function of the potential will have tails that decay steeply, even
steeper than those of the normal distribution.

\subsection{Large and small potentials. The saddle point approximation}
\begin{figure}
\includegraphics{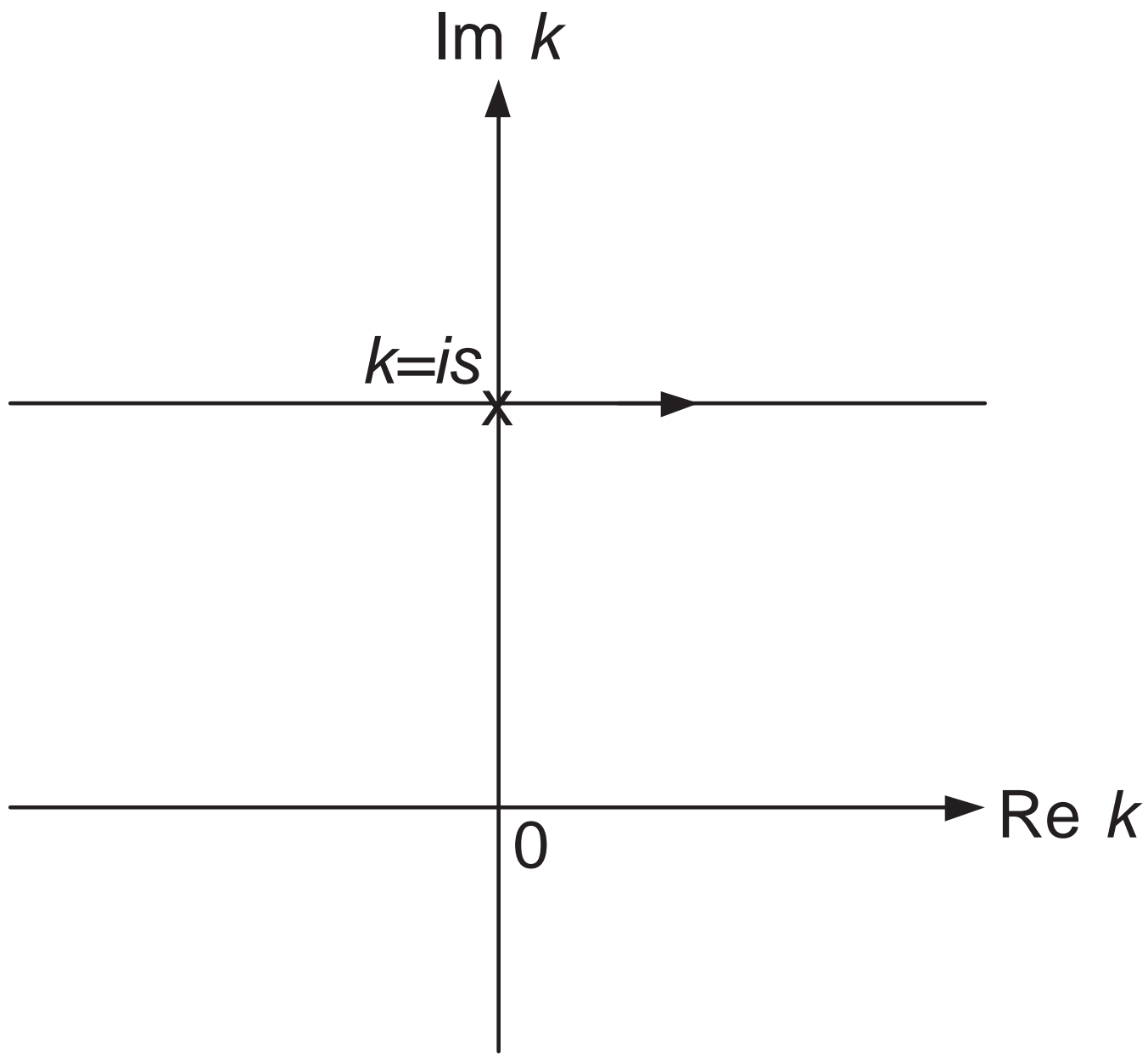}
\caption{The integration contour passes through the saddle point
$k=is$ in the complex plane.}\label{f:contour}
\end{figure}

The existence of the first moment of the sum (\ref{eq:ber})
depends on its tail distribution, which we calculate below by the
saddle point method \cite{Jensen}. That is, we calculate the
chance of finding a pinch of (noncrystalline) salt with a very
large potential. For a potential $\Phi$ defined in equation
(\ref{eq:ber}), we denote the pdf of
$\left(\ds\frac{q}{4\pi\varepsilon_0 R} \right)^{-1}\Phi$ by
$f(x)$. The Fourier transform $\hat{f}(k)$ of this pdf is given by
the infinite product
\begin{equation}
\hat{f}(k) = \prod_{n=0}^\infty \cos\left(\frac{k}{1+na} \right),
\end{equation}
which is an entire function in the complex plane, because the
general term is $1+O(n^{-2})$. The inverse Fourier transform
recovers the pdf
\begin{equation}
f(x) = \frac{1}{2\pi} \ds \int_{-\infty}^\infty \hat{f}(k)
e^{ikx}\,dk,
\end{equation}
which we want to evaluate asymptotically for large $x$. Setting
\begin{equation}
\label{eq:g} g(k,x) = \sum_{n=0}^\infty \log
\cos\left(\frac{k}{1+na} \right) + ikx,
\end{equation}
we write
\begin{equation}
\label{eq:Fourier} f(x) = \frac{1}{2\pi} \ds \int_{-\infty}^\infty
\exp\{g(k,x)\}\,dk.
\end{equation}
The saddle point is the point $k$ for which $\ds
\frac{d}{dk}g(k,x)=0$. Differentiating equation (\ref{eq:g}) with
respect to $k$, we find that
\begin{equation}
\frac{d}{dk}g(k,x) = -\sum_{n=0}^\infty
\frac{\tan\left(\ds\frac{k}{1+na} \right)}{1+na} + ix.
\end{equation}
We look for a root of the derivative on the imaginary axis, and
substitute $k=is$. The vanishing derivative condition of the
saddle point method is then
\begin{equation}
\label{eq:saddle} x = \sum_{n=0}^\infty
\frac{\tanh\left(\ds\frac{s}{1+na} \right)}{1+na}.
\end{equation}
The infinite sum on the right hand side represents a monotone
increasing function of $s$ in the interval $0<s<\infty$, so
equation (\ref{eq:saddle}) has exactly one solution for every $x$.
Near the saddle point $k=is$, we approximate $g(k)$ by its Taylor
expansion up to the order
\begin{equation}
\label{eq:Taylor} g(k) \approx g(is) +
\frac{1}{2}\frac{d^2}{dk^2}g(is)(k-is)^2,
\end{equation}
to find the leading order term of the full asymptotic expansion
(derivatives of higher order of the Taylor expansion can be used
to find all terms of the asymptotic expansion \cite{Carrier}). We
use the Cauchy integral formula to calculate our Fourier integral
(\ref{eq:Fourier}) on the line parallel to the real $k$ axis
through $k=is$ (see Fig. \ref{f:contour})
\begin{eqnarray}
f(x) &\approx& \frac{1}{2\pi} e^{g(is)}\int_{-\infty}^\infty
\exp\left\{\frac{1}{2}g''(is)(k-is)^2\right\}\,dk \nonumber \\
&&\nonumber\\ &=& \frac{1}{2\pi} e^{g(is)} \int_{-\infty}^\infty
\exp\left\{ g''(is)\frac{z^2}{2}\right\}\,dz =
\frac{e^{g(is)}}{\sqrt{-2\pi g''(is)}}. \label{eq:saddle-approx}
\end{eqnarray}
Equation (\ref{eq:saddle}) has no analytic solution, so we
construct asymptotic approximations for large and small values of
$s$ separately.

\subsection{Tail asymptotics}
Throughout this subsection we assume that $a$ is small and $s$ is
large and we find the tail asymptotics of the pdf away from the
system (for $a\ll1$). For $s\gg 1$ the Euler-Maclaurin sum formula
gives
\begin{equation}
x = \int_0^\infty \frac{\tanh\left(\ds\frac{s}{1+ax}
\right)}{1+ax}\,dx+\frac{1}{2}\tanh s+O(a).
\end{equation}
Substituting $z=\ds\frac{s}{1+ax}$, we obtain
\begin{equation}
\label{eq:EM} x = \frac{1}{a}\int_0^s \frac{\tanh
z}{z}\,dz+\frac{1}{2}\tanh(s)+O(a).
\end{equation}
Writing
\begin{eqnarray}
\int_0^s \frac{\tanh z}{z}\,dz &=& \int_0^1 \frac{\tanh z}{z}\,dz
+ \int_1^s \frac{\tanh z-1}{z}\,dz + \int_1^s \frac{dz}{z}
\nonumber \\
&=& \log s +\int_0^1 \frac{\tanh z}{z}\,dz + \int_1^\infty
\frac{\tanh z-1}{z}\,dz +  O(e^{-2s}), \nonumber
\end{eqnarray}
we obtain (\ref{eq:EM}) in the form
\begin{equation}
\label{eq:logs} ax = \log s + C +\frac{a}{2}+O(a^2,e^{-2s}),
\end{equation}
where the constant $C$ is given by
\begin{equation}
\label{eq:C} C=\int_0^1 \frac{\tanh z}{z}\,dz + \int_1^\infty
\frac{\tanh z-1}{z}\,dz.
\end{equation}
Exponentiation of equation (\ref{eq:logs}) gives the location of
the saddle point asymptotically for small $a$ and large $s$ as
\begin{equation}
\label{eq:s-exp} s = e^{ax-C-a/2+O(a^2,e^{-2s})}.
\end{equation}
The saddle point approximation (\ref{eq:saddle-approx}) requires
the evaluation of $g$ and its second derivative at $k=is$. The
Euler-Maclaurin sum formula gives
\begin{eqnarray}
g(is) &=& \sum_{n=0}^\infty \log
\cosh\left(\frac{s}{1+na} \right)-sx \nonumber \\
&&\nonumber\\
 &=& \frac{s}{a}\int_0^s \frac{\log \cosh z}{z^2}\,dz
+ \frac{1}{2}\log \cosh s -sx + O(as)
\nonumber \\
&&\nonumber\\ &=& \frac{s}{a} \left(\int_0^1 \frac{\log \cosh
z}{z^2}\,dz + \int_1^s \frac{dz}{z} + \int_1^\infty \frac{\log
\cosh z - z}{z^2}\,dz + O\left(\frac{1}{s}\right)
\right) \nonumber \\
&&\nonumber\\ & & + \frac{s}{2} - \frac{\log 2}{2} -sx + O(as).
\nonumber
\end{eqnarray}
Using equations (\ref{eq:logs}) and (\ref{eq:C}), we find
\begin{equation}
g(is) = C_1\,\frac{s}{a} - \frac{\log 2}{2} +
O\left(a,\frac{1}{a},as\right),
\end{equation}
where
\begin{equation}
C_1 = \int_0^1 \frac{\log \cosh z}{z^2}\,dz + \int_1^\infty
\frac{\log \cosh z - z}{z^2}\,dz - \int_0^1 \frac{\tanh z}{z}\,dz
- \int_1^\infty \frac{\tanh z-1}{z}\,dz,
\end{equation}
and integration by parts shows that $C_1=-1$. It follows that
\begin{equation}
\label{eq:g0} g(is) = - \frac{s}{a} - \frac{\log 2}{2} +
O\left(a,\frac{1}{a},as\right).
\end{equation}
The second derivative of $g$ is evaluated in a similar fashion
\begin{eqnarray}
\frac{d^2}{dk^2}g(k)\bigg|_{k=is} &=& -\sum_{n=0}^\infty
\ds\frac{1-\tanh^2\left(\ds\frac{s}{1+na} \right)}{(1+na)^2}
\nonumber\\
&&\nonumber\\
&=& -\frac{1}{as} \int_0^{s}\left(1-\tanh^2 z
\right)\,dz -
\frac{1}{2}\left(1-\tanh^2 s \right)+O(ase^{-2s}) \nonumber \\
&&\nonumber\\
&=& -\frac{\tanh s}{as} + O(ase^{-2s}, e^{-2s}) \nonumber \\
&&\nonumber\\ &=& -\frac{1}{as} + O(as,1,\frac{1}{as})e^{-2s}.
\label{eq:g''}
\end{eqnarray}
Substitution of (\ref{eq:g0}), (\ref{eq:g''}), and
(\ref{eq:s-exp}) into the saddle point approximation
(\ref{eq:saddle-approx}) gives
\begin{eqnarray}
f(x) &\approx& \frac{\sqrt{a}}{2\sqrt{\pi}}
e^{\frac{1}{2}\left(ax-C-a/2 \right)}
e^{-\frac{1}{a}e^{ax-C-a/2}},
\end{eqnarray}
where the constant $C=.8187801402\cdots$ is given by equation
(\ref{eq:C}). Therefore, the small $a$ and large $s$ approximation
to the tail of the pdf of $\Phi$ is given by
\beq
&& f_\Phi(x) \sim\label{eq:pdf-phi}\\
&&\nonumber\\
&& \frac{4\pi\varepsilon_0 R}{q} \frac{\sqrt{a}}{2\sqrt{\pi}}
\exp\left\{{\ds\frac{1}{2}\left(\ds\frac{4\pi\varepsilon_0
d}{q}x-C-a/2
\right)-\ds\frac{1}{a}\,\exp\left\{\ds\frac{4\pi\varepsilon_0
d}{q}x-C-a/2\right\}}\right\}, \quad x\to\infty.\nonumber \eeq It
follows from equation (\ref{eq:pdf-phi}) that the pdf decays to
zero as a double exponential as $x\to \infty$, which implies that
all moments exist. This decay is similar to the extreme value or
the log-Weibull (Gumbel) distributions \cite{Weibull}. The compact
support of the distributions of convergent series is replaced here
with a steep decay. Note also that the decay becomes steeper
further away from the system, as expected, because the
pre-exponential factor of the inner exponent is $1/a=R/d$.

For small $x$ the pdf can be approximated by a zero mean Gaussian
with variance $Var\,(\Phi)$, which for small $a$ is
\begin{equation}
f_\Phi(x) \sim
\frac{4\pi\varepsilon_0}{q}\sqrt{\frac{Rd}{2\pi}}\exp\left\{-\ds\frac{Rd}{2}\left(
\ds\frac{4\pi\varepsilon_0 x}{q}\right)^2\right\}, \quad x\to 0.
\end{equation}
Near its center, the distribution looks like a Gaussian with a
standard deviation that decays like $1/\sqrt{R}$, in agreement
with equation (\ref{eq:sd}). We conclude that the pdf looks normal
near its center, but, far away from there, it decays to zero much
more steeply, rather like a cutoff. This conclusion is the answer
to the question posed in subsection \ref{sec:weighted} about the
normality of weighted sums of i.i.d. random variables. The non
Gaussian tails of the distribution are characteristic of large
deviations \cite{Jensen}.

\section{Random Distances}
Consider a one-dimensional system of alternating charges without
the restriction of equal distance between successive charges. In
particular, we assume a renewal model, in which the distances
between two neighboring charges are non-negative i.i.d random
variables with pdf $f(l)$ and finite expectation value
$$d = \int_0^\infty lf(l)\,d\l < \infty.$$
The potential of this random system is also a random variable.

We show below that away from the system the mean value of the
potential $\bar{V}$ has the asymptotic form
\begin{equation}
\label{random-screening} \bar{V} =
\frac{q}{4\pi\varepsilon_0R}\left(\frac12+O(a) \right),
\end{equation}
where $a=d/R$. Equation (\ref{random-screening}) defines the
attenuation produced by the configuration of charges. The mean
potential of the system is produced by (in effect) half a charge.
We note that the value $1/2$ is exactly the same for both random
and non-random systems of alternating charges
(eq.(\ref{eq:asym})). We first note that
\begin{equation}
\label{eq:renewal} \Pr\{V(R)=V\} = \int_0^\infty f(l)
\Pr\left\{V(R+l)=\frac{q}{4\pi\varepsilon_0R}-V\right\}\,dl.
\end{equation}
To find the mean value, we multiply (\ref{eq:renewal}) by $V$ and
integrate (note that $0\leq V \leq
\displaystyle\frac{q}{4\pi\varepsilon_0R}$), and then change the
order of integration
\begin{eqnarray}
\bar{V}(R) &=&
\int_0^{\displaystyle\frac{q}{4\pi\varepsilon_0R}}V\,dV
\int_0^\infty
f(l)\Pr\left\{V(R+l)=\frac{q}{4\pi\varepsilon_0R}-V\right\}\,dl
\nonumber \\
&&\nonumber\\
 &=& \int_0^\infty f(l)\,dl
\int_0^{\displaystyle\frac{q}{4\pi\varepsilon_0R}}V\Pr\left\{V(R+l)=
\frac{q}{4\pi\varepsilon_0R}-V\right\}\,dV \nonumber \\
&&\nonumber\\
&=& \frac{q}{4\pi\varepsilon_0R} - \int_0^\infty
f(l)\,dl
\int_0^{\displaystyle\frac{q}{4\pi\varepsilon_0R}}V\Pr\{V(R+l)=V\}\,dV
\nonumber \\
&&\nonumber\\ &=& \frac{q}{4\pi\varepsilon_0R} - \int_0^\infty
f(l)\bar{V}(R+l)\,dl. \label{eq:barV}
\end{eqnarray}
We look for an asymptotic expansion of the form
\begin{equation}
\bar{V}(R) = \frac{q}{4\pi\varepsilon_0} \left(\bar{V}_0+a
\bar{V}_1 + a^2 \bar{V}_2 + \ldots \right).
\end{equation}
Substituting this asymptotic expansion into (\ref{eq:barV}) gives
$\bar V_0=1/2$ for the $O(1)$ term, because
\begin{equation}
1-a \leq \int_0^\infty f(l) \frac{R}{R+l}\,dl \leq 1.
\end{equation}
The first inequality is due to the inequality $\displaystyle
\frac{1}{1+x} \geq 1-x$. Hence (\ref{random-screening}) follows.

\section{Dimensions Higher than One}
\subsection{The condition of global electroneutrality}
In dimensions higher than one, global electroneutrality is enough
to dramatically attenuate the electric field, but it is not enough
to produce a small potential, as shown below.

Consider the electric potential at a vacant site of random charges
located at the points of a 2D square lattice
\begin{equation}
\Phi = \sum_{(n,m)\neq(0,0)} \frac{X_{nm}}{\sqrt{n^2+m^2}}.
\end{equation}
The variance of $\Phi$ is
\begin{equation}
Var( \Phi) = \sum_{(n,m)\neq(0,0)} \frac{1}{n^2+m^2} = \infty.
\end{equation}
The infinite value of the variance means that arbitrarily large
potentials can occur with high probability. That is, the electric
potential is not attenuated. The divergence of the variance of the
potential of three dimensional systems is even steeper. Therefore,
attenuation of the potential of totally disordered systems  can
occur in two or three dimensional systems \emph{only} if some
correlation is introduced into the distribution of the locations of
the charges. If, for example, the signs of all charges alternate, as
in a real Na$^+$Cl$^-$ crystal, the distribution of potential will
be dramatically different, and greatly attenuated, compared to a two
or three dimensional system in which many charges of one sign are
clumped together.

The condition of global electroneutrality is enough to ensure the
dramatic attenuation of the electric field. Indeed, consider a 3D
cubic lattice of random charges. The $z$-component of the electric
field at a vacant lattice point is
\begin{equation}
E_z = \sum_{(n,m,l)\neq (0,0,0)} \frac{X_{nml}\cos \left(
\ds\frac{n}{\sqrt{n^2+m^2+l^2}}\right)}{n^2+m^2+l^2}.
\end{equation}
The variance of $E_z$ is finite,
 \beqq
Var(E_z) = \sum_{(n,m,l)\neq (0,0,0)} \frac{\cos^2 \left(
\ds\frac{n}{\sqrt{n^2+m^2+l^2}}\right)}{(n^2+m^2+l^2)^2} < \infty,
 \eeqq
because convergence is determined by the integral
 $$2\pi\int_0^\pi \cos^2\theta \sin\theta \,d\theta \int_d^\infty
 \frac{1}{r^4}r^2\,dr < \infty.$$
The large potential means that much work has to be done to create
the given spatial configuration of the charges, however, the
resulting field remains usually small.

\subsection{The condition of local electroneutrality}
\begin{figure}
\includegraphics{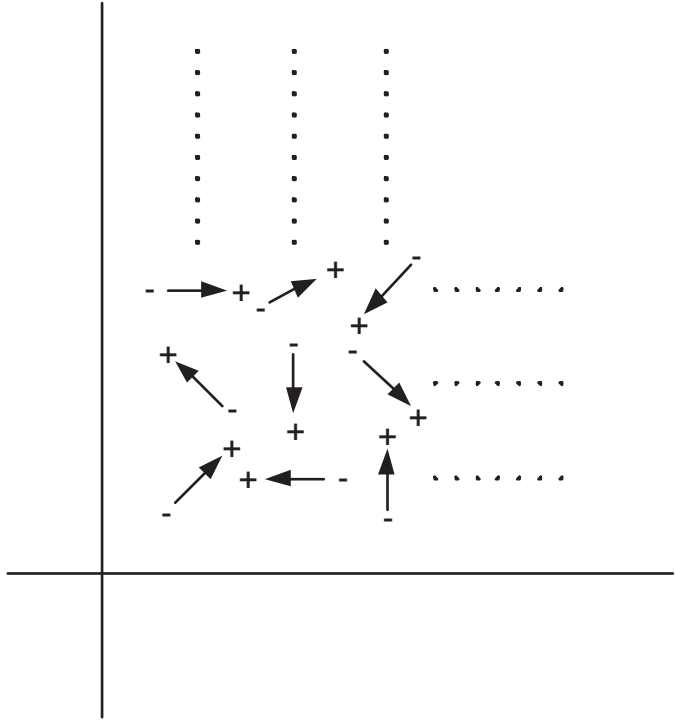}
\caption{Two dimensional lattice of dipoles of randomly chosen
orientations produce attenuation due to the condition of local
electroneutrality.}\label{f:dipoles}
\end{figure}

Here we show that the condition of local electroneutrality implies
the attenuation of the potential in two and three dimensions. For
example, the potential of a two or three dimensional lattice of
extended dipoles is finite with probability 1, if the orientation
of dipoles is distributed independently, identically, and
uniformly on the unit sphere (see Fig. \ref{f:dipoles}).

Paraphrasing \cite[p.136]{Jackson}, we say that a (net) charge
distribution $\rho(\x)$ has {\em local charge neutrality} if the
(net) charge inside a sphere of radius $R$ falls with increasing
$R$ faster than any power, that is, for any $\x$
\begin{equation}
\label{eq:local} \lim_{R\to\infty}R^n\int_{|\x-\y|<R}
\rho(\y)\,d\y = 0\quad\mbox{for all $n>0$}.
\end{equation}
On a lattice, the number of charges that are assigned to each
lattice point can be larger than in our example of dipoles (Fig.
\ref{f:dipoles}), thus forming multipoles. The Debye-H\"uckel
distribution also satisfies the local charge neutrality condition.

The potential of a single lattice point can then be written as an
expansion in spherical harmonics, if the charges of each multipole
are contained in a single lattice box. It can be also expanded, if
the charge density of each multipole decays sufficiently fast, as
\cite{Jackson}
\begin{equation}
\Phi_{0,0,0}(\mb{x}) =\frac{1}{4\pi\varepsilon_0}
\sum_{l=0}^\infty \sum_{m=-l}^l
\frac{1}{2l+1}\,q_{lm}\,\frac{Y_{lm}(\theta,\phi)}{r^{l+1}},
\end{equation}
where $q_{lm}$ are the multipole moments. In particular, the
zeroth order multipole moment is
\begin{equation}
q_{00} = \frac{1}{\sqrt{4\pi}} \int \rho(\y)\,d\y = 0,
\end{equation}
by the condition of local electroneutrality (\ref{eq:local}): the
far potential due to a single lattice point decays as $1/r^2$ (or
steeper). The coefficients $q_{lm}$ assigned to each lattice point
are randomized as in the previous sections so their mean value
vanishes, meaning that there is no preferred orientation in space.
(Compare the example of dipoles which do not have a preferred
orientation.) The mean value of the potential of the entire
lattice is then $\langle\Phi\rangle=0$. The variance is given by
\begin{equation}
\label{eq:3D-var} Var(\Phi) = \sum_{ijk} Var(\Phi_{ijk}),
\end{equation}
where $\Phi_{ijk}$ is the potential of the charge at lattice point
$(i,j,k)$. The potential decays as $1/r^2$ (or steeper); therefore
the variance decays as $1/r^4=1/(i^2+j^2+k^2)^2$ (or steeper). The
convergence of the infinite sum (\ref{eq:3D-var}) is determined by
the convergence of the integral
\begin{equation}
\int_{r>d} \frac{1}{r^4}\,dV = 4\pi \int_d^\infty
\frac{1}{r^2}\,dr = \frac{4\pi}{d} < \infty.
\end{equation}
Thus, the variance of the potential is finite and we have shown
that local electroneutrality produces a dramatic attenuation of
potential. As above, the potential away from a charge is usually
of the order of the potential of a single charge.

\subsection{The liquid state}
Screening in the liquid state involves a least three phenomena. (1)
The movement of charge to a distribution of minimal free energy. (2)
The properties of a static charge distribution with minimal free
energy. (3) The properties of any charge distribution.

If the charge correlation function $\rho(\x)$ minimizes free
energy, and is at equilibrium, as in ionic solutions, the far
field potential is strongly screened. However, the relaxation into
such a state takes time, typically psec to nsec in an ionic
solution under biological conditions (see measurements reported in
\cite{Barthel2}, and theory summarized in \cite{Knodler}). As long
as local charge neutrality exists during the relaxation period,
the potential changes from attenuated (as described above) to
exponentially screened, as equilibrium is reached. In fact, the
spread of potential in ionic solutions has the curious property
that it is much less shielded at short times than at long times;
potentials on the (sub) femtosecond time scale of atomic dynamics
spread macroscopic distances while potentials on long time scales
spread only atomic distances. Specifically, potentials on a time
scale greater than nano or microseconds spread a few Debye
lengths, only a nanometer or so under biological conditions,
although potentials on a femtosecond time scale can spread
arbitrarily far depending on the configuration of dielectrics at
boundaries that govern the violations of local electroneutrality.
To make this verbal analysis of fast phenomena rigorous, the
potentials and fields should be computed from Maxwell's equations,
not Coulomb's law.

Non-equilibrium fluctuations may violate local charge neutrality,
therefore field fluctuations can be large. For example, in systems
which are not locally electroneutral, potential can spread a long
way, as in the telegraph \cite{Ghausi}, Kelvin's transatlantic
cable, or the axons of nerve cells \cite{Jack}. In such systems,
d.c. potential spreads arbitrarily far---kilometers in telegraphs;
thousands of kilometers in the transatlantic cable; centimeters in
a squid nerve filled with salt water---even if an abundance of
ions ($\approx 10^{23}$) are present. Local electroneutrality is
violated in such systems (at the insulating boundary which
separates the inside and outside of the cable, e.g., the cell
membrane) and that violation allows large far field potentials.

\section{Summary and Discussion}
Global electroneutrality ensures the dramatic attenuation of the
electric potential and field of a one dimensional system of
charges. Even if local electroneutrality is violated, and the
local net charge is not zero, the potential remains finite in
these one dimensional systems, even in a  random lattice that
includes arbitrarily long strings of equal charges. We have shown
that the distribution of the weighted sum of i.i.d. random
variables that define the one-dimensional electric potential is
almost normal near its center, but has very steep double
exponentially decaying tails. The distances between neighboring
charges can also be random, without changing the attenuation
effect. In higher dimensions, global electroneutrality is
sufficient to dramatically attenuate the electric field, but not
the potential. However, local electroneutrality ensures a small
potential in two and three dimensions, so the electric potential
and field is short range in one, two, and three dimensions, if the
systems are \emph{locally} electroneutral.

\section{Acknowledgement}
The comments of David Ferry, Mark Ratner,  Stuart Rice,\textbf{
and the referees }were most helpful. This research was partially
supported by research grants from the Israel Science Foundation,
US-Israel Binational Science Foundation, and the NIH Grant No.
UPSHS 5 RO1 GM 067241.

\newpage

\newpage

\newpage

\end{document}